\documentclass[useAMS,usenatbib]{mn2e}
\usepackage{epsfig}
\def\lesssim{\mathrel{\hbox{\rlap{\hbox{\lower4pt\hbox{$\sim$}}}\hbox{$<$}}}}
\def\gtrsim{\mathrel{\hbox{\rlap{\hbox{\lower4pt\hbox{$\sim$}}}\hbox{$>$}}}}

\def\aap{A\&\hskip-1pt A}

\title[Open clusters NGC 1245 and NGC 2506]
{Deep and Wide Photometry of Two Open Clusters NGC 1245 and NGC
2506: CCD Observation and Physical Properties}

\author[Lee, Kang \& Ann]{S. H. Lee,$^1$\thanks{E-mail:ngc2420@hanmail.net} 
Y.-W. Kang$^1$ and H. B. Ann$^2$\thanks{Author to whom any
correspondence should be addressed. E-mail:hbann@pusan.ac.kr}\\
$^1$Korea Astronomy and Space Science Institute, Daejeon 305-348, Korea\\
$^2$Department of Earth Sciences, Pusan National University, Busan 609-735, Korea}

\begin{document}

\date{Accepted 2012 June 26. Received 2012 June 26; in original form 2012 March 1}

\pagerange{\pageref{firstpage}--\pageref{lastpage}} \pubyear{2012}

\maketitle

\label{firstpage}

\begin{abstract}

We have conducted $VI$ CCD photometry of the two open clusters NGC
1245 and NGC 2506 using the CFH12K CCD camera. Our photometry covers
a sky area of $84^\prime \times 82^\prime$ and $42^\prime \times
81^\prime$ for the two clusters, respectively, and reaches down to
$V \approx 23$. We derived the physical parameters using detailed 
theoretical isochrone fittings using $\chi^2$ minimization. The derived 
cluster parameters are $E(B-V) = 0.24 \pm0.05$ and 
$ 0.03 \pm0.04$, $(V-M_V)_0  = 12.25 \pm0.12$ and $12.47 \pm0.08$,
$age(Gyr) = 1.08 \pm0.09$ and $2.31 \pm0.16$, and [Fe/H]= $-0.08 \pm0.06$
and $-0.24 \pm0.06$, respectively for NGC 1245 and NGC 2506, 
We present the luminosity functions (LFs) of the two clusters, 
which reach down to $M_V \approx 10$, 
and derive mass functions (MFs) with slopes of $ \Gamma = -1.29$ for 
NGC 1245 and $ \Gamma = -1.26$ for NGC 2506. The slopes are slightly
shallower than that of the solar neighbourhood, implying the
existence of dynamical evolution that drives the evaporation of the
low-mass stars in the clusters.

\end{abstract}


\begin{figure*}
 \epsfig{figure=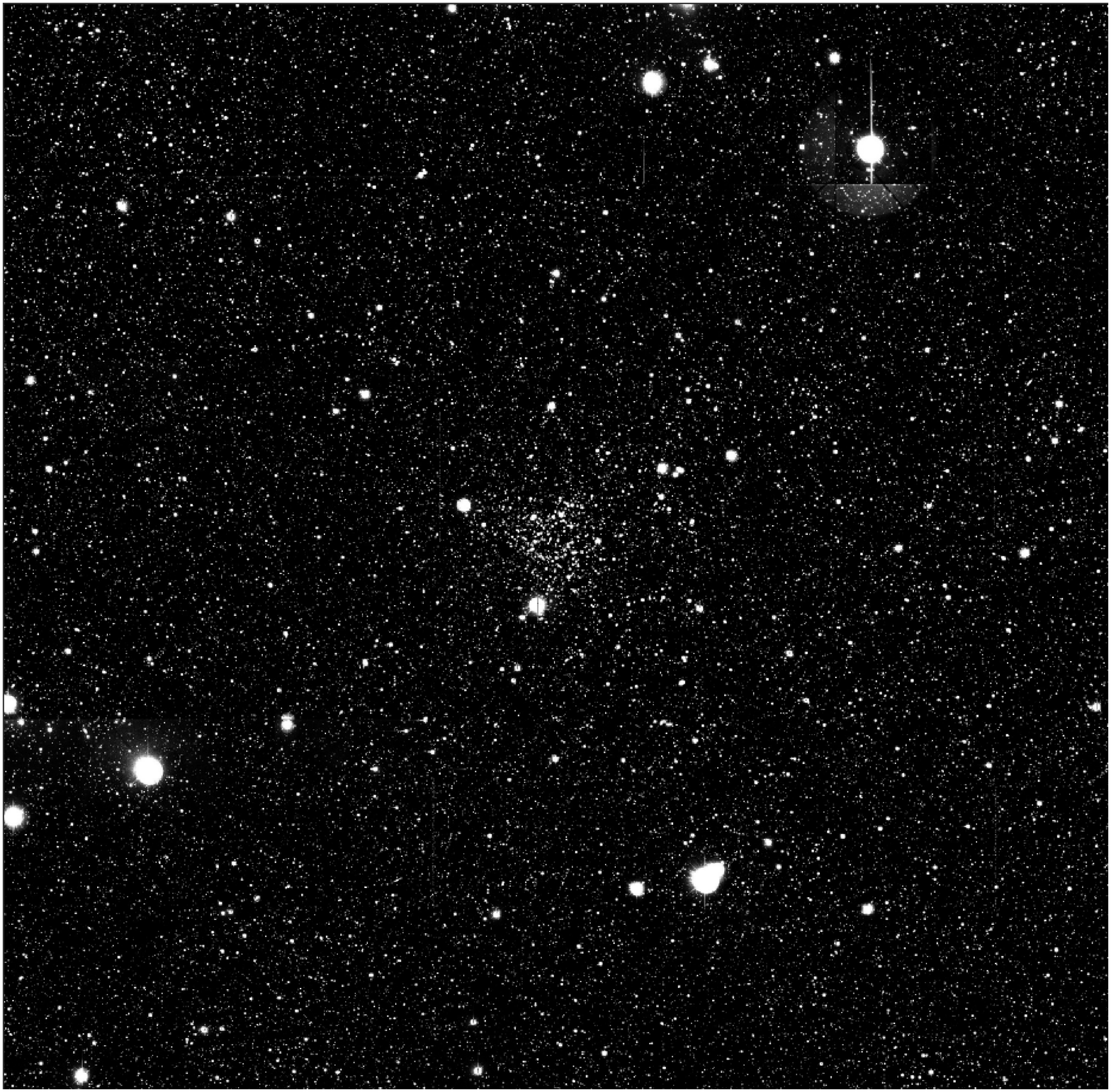, height=0.96\textwidth, width=0.96\textwidth}
 \vspace{2mm}
 \caption{Observed fields ($84^{\prime} \times 82^{\prime}$) of NGC 1245.
We combined 6 V-filter images, where one CCD image has a size of
$42^{\prime} \times 28^{\prime}$, to make the mosaic image. North is
up and east is to the left. }
 \vspace{-1mm}
 \label{fig_fov1}
\end{figure*}

\section{Introduction} 

Open clusters are located in the Galactic disc as Population  I, 
and it is important to investigate their properties and the spatial
distribution of the Galaxy to understand the Galaxy structure and
evolution, especially the formation and dynamical evolution of the
Galactic disc.  
Because stars in the Galactic disc are originated in evaporated stars from 
open clusters, the 
luminosity function of open clusters is an
important parameter for understanding their formation and evolution.
We can easily obtain the LF from CCD observations, but an analysis
of the LF is not easy because the observed LF of open clusters reflects
the dynamical evolution which depends on the cluster's age as well as 
the initial mass function (IMF). 

It is necessary to have a deep photometry over the entire cluster field
to understand the LF of old open clusters down to the magnitudes of the
evaporating stars because they are much fainter than the turn-off stars and
are supposed to be located around the tidal radius of the clusters.
But there are only a few open
clusters of which photometry is deep and wide enough to study the effect of 
dynamical evolution on the structure of open clusters. Thus, we selected
open clusters which are old enough for the low mass stars to evaporate
from the clusters by dynamical evolution and rich enough suitable for a 
detailed analysis of the spatial distribution of low mass stars. We also
took into account the distance of open clusters to ensure that our photometry
reaches up to $M_V \sim 10$.

The old open cluster NGC 1245 is located in the Perseus arm at a
Galactic latitude of $-8^{\circ}.93$. Using photographic photometry
data from the Schmidt telescope, \citet{per78} discussed the radial
density distribution of stars with a limiting magnitude of $B =
17$ and a radius of $r \approx 83^{\prime}$. They presented cluster
boundaries with a nucleus ($r < 13^{\prime}$) and corona ($
13^{\prime} < r < 61^{\prime}$). \citet{car94} and \citet{sub03}
found a shallower slope of the mass function of NGC 1245 than
\citet{sal55}. Using $BV$ CCD photometry, \citet{sub03} presented
the surface number density with a limiting magnitude of $V = 17$
and a radius of $r \sim 6^{\prime}.5$. They also indicated a lack of
stars in the cluster centre. \citet{bur04} derived physical
parameters and a core radius of $R_c = 3^{\prime}.1$ from $BVI$ CCD
photometry. They determined an optical absorption of $A_V = 0.68$
magnitude and a distance modulus $(V-M_V)_0 = 12.27$; they also reported
that the cluster has no differential reddening, in contrast to
previous studies such as \citet{car94} and \citet{wee96}.

The old open cluster NGC 2506 also has rich members located at a
Galactic latitude of $+9^{\circ}.94$. A radial density profile of
the proper motion membership stars obtained by \citet{chi81}
indicated that bright stars are more concentrated than faint stars
in this cluster. \citet{mar97} suggested that more than 20\% of the
main sequence stars are binary. From a photometric study by
\citet{hen07}, the distance of this cluster is known to be 3.4 kpc
and its age is 1.79 Gyr.

This paper presents the deep $VI$ CCD photometry of the two 
old open clusters NGC 1245 and NGC 2506, covering the entire cluster fields 
including the surrounding field regions for field star correction,
along with the derivation of the cluster parameters such as age, metallicity,
distance, and interstellar reddening. We will discuss the dynamical
structures of the open clusters in a forthcoming paper
(Lee, Kang $\&$ Ann, in preparation), but here we present the physical properties,
luminosity functions, and mass functions of the two open clusters
along with their physical properties.

The present paper is organized as follows. In \S2, we describe the
observations and data reductions. We present the physical parameters
of the clusters in \S3, and describe the luminosity functions and
mass functions in \S4 and \S5, respectively. A summary is given in the final section.

\begin{figure}
 \vspace{2mm}
 \epsfig{figure=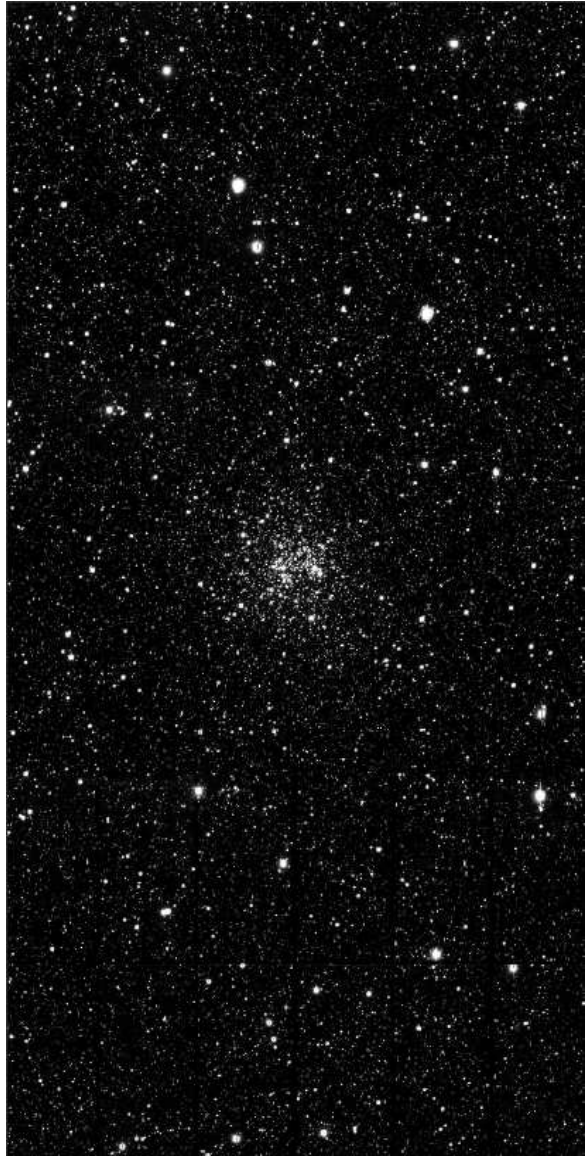, height=0.96\textwidth, width=0.48\textwidth}
 \vspace{2mm}
 \caption{
Observed fields ($42^{\prime} \times 81^{\prime}$) of NGC 2506.
We combined 3 V-filter images, where one CCD image has a size of
$42^{\prime} \times 28^{\prime}$, to make the mosaic image. North is
up and east is to the left. }
 \vspace{-2mm}
 \label{fig_fov3}
\end{figure}

\section{Observation}

We selected target clusters best suited for studying the dynamical properties
of open clusters which is the subject of the second paper \citep{lee12}.
The two target clusters NGC 1245 and NGC 2506
are located at a relatively high Galactic latitude to reduce field star 
contamination and are old enough for the evaporation of low mass stars to 
be prevalent due to dynamical evolution. 
Also, they are at proper distances for optimum sky coverage and the deep 
photometry to reach up to $M_{V}\sim 10$. 
The basic known properties of the selected clusters are listed in Table 1.

\begin{table*}
 \centering
 \begin{minipage}{125mm}
 \raggedright
  \caption{Basic data of the selected clusters, extracted from the WEBDA web site (http://obswww.unige.ch/webda/)}
   \begin{tabular}{@{}cccccccc@{}} \hline
      Cluster    &      $\alpha$                  &  $\delta$  &             $l$          &           $b$           &   Distance  &$E(B-V)$& log(age)   \\
       name     &       J2000.0                     &   J2000.0   &                           &                          &     (kpc)     &            &               \\ \hline\hline
    NGC 1245  &  $03^{h} 14^{m} 42^{s}$   &  +47 14 12  & $146^{\circ}.65$  & $-8^{\circ}.93$  &      2.88      &  0.300   &  8.704   \\
    NGC 2506  &  $08^{h} 00^{m} 01^{s}$   &  -10 46 12  & $230^{\circ}.56$  & $+9^{\circ}.94$   &     3.46      &  0.081   &  9.045   \\
   \hline
  \end{tabular}
 \end{minipage}
 \vspace{5pt}
\end{table*}

\subsection{Observation}

We conducted deep $VI$ CCD photometry using the CFH12K mosaic CCD
camera mounted at the prime focus of the Canada-France-Hawaii
Telescope (CFHT). The effective aperture of the CFHT is 3.6 m, and
the CFH12K mosaic CCD camera consists of 12 CCDs of $2048\times4096$
pixels. The CFH12K field of view is $42^{\prime}\times28^{\prime}$
with pixel size of $0.^{\prime\prime}206$.

For better sky coverage, we observed 6 regions for NGC 1245 and 3
regions for NGC 2506. The field of view of the combined images was
$84^{\prime} \times 82^{\prime}$ for NGC 1245 and $42^{\prime}
\times 81^{\prime}$ for NGC 2506 (see Fig. 1 and Fig. 2). The
observations were performed on the night of December 6, 2001, at
CFHT. The observation log is given in Table 2.

\begin{table*}
 \centering
 \begin{minipage}{95mm}
 \raggedright
  \caption{Log of observations}
   \begin{tabular}{@{}ccccc@{}} \hline
      Cluster    &     Region        &             $V$ Filter                    &                 $I$  Filter               &   Seeing    \\ \hline\hline
    NGC 1245  &  North-East   &  $140^{s} \times$ 4, $5^{s} \times$ 4  &  $90^{s} \times$ 4, $5^{s} \times$ 4  &  $1.^{\prime\prime}6$  \\
                    &  East            &  $140^{s} \times$ 4, $5^{s} \times$ 4  &  $90^{s} \times$ 4, $5^{s} \times$ 4  &  $1.^{\prime\prime}2$  \\
                    &  South-East   &  $140^{s} \times$ 4, $5^{s} \times$ 4  &  $90^{s} \times$ 4, $5^{s} \times$ 4  &  $1.^{\prime\prime}1$  \\
                    &  North-West   &  $140^{s} \times$ 4, $5^{s} \times$ 4  &  $90^{s} \times$ 4, $5^{s} \times$ 4  &  $1.^{\prime\prime}3$  \\
                    &  West            &  $140^{s} \times$ 4, $5^{s} \times$ 4  &  $90^{s} \times$ 4, $5^{s} \times$ 4  &  $1.^{\prime\prime}5$  \\
                    &  South-West   &  $140^{s} \times$ 4, $5^{s} \times$ 4  &  $90^{s} \times$ 4, $5^{s} \times$ 4  &  $1.^{\prime\prime}3$  \\
    NGC 2506  &  North            &  $140^{s} \times$ 3, $5^{s} \times$ 3  &  $90^{s} \times$ 4, $5^{s} \times$ 3  &  $1.^{\prime\prime}2$  \\
                    &  Centre          &  $140^{s} \times$ 3, $5^{s} \times$ 3  &  $90^{s} \times$ 4, $5^{s} \times$ 3  &  $1.^{\prime\prime}0$  \\
                    &  South           &  $140^{s} \times$ 3                       &  $90^{s} \times$ 4                      &  $1.^{\prime\prime}2$  \\
   \hline
  \vspace{-1mm}
  \end{tabular}
 \end{minipage}
\end{table*}


\begin{figure}
 \vspace{-11mm}
 \epsfig{figure=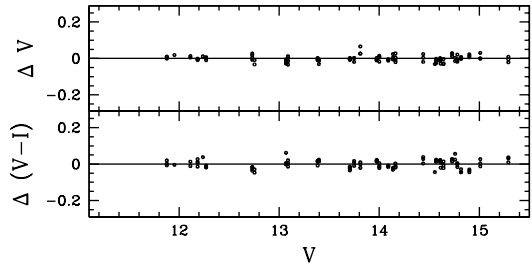, height=0.480\textwidth, width=0.480\textwidth}
 \vspace{-25mm}
 \caption{
   Photometric residuals of the transformation to the standard system. 
  We used standard stars in SA 98 \citep{lan92}.
Standard deviations of the $VI$ photometry are $\sigma_{\Delta V} = 0.018$ and
 $\sigma_{\Delta(V-I)} = 0.028$.
}
 \vspace{-2mm}
 \label{std_res}
\end{figure}

\begin{figure}
 \vspace{-2pt}
 \epsfig{figure=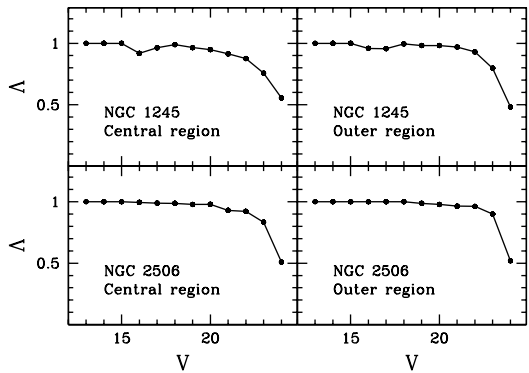, height=0.48\textwidth, width=0.48\textwidth}
 \vspace{-22mm}
 \caption{
 Incompleteness correction factor $\Lambda$ as a function of $V$ magnitude.
}
 \vspace{-4mm}
 \label{complet_cor}
\end{figure}

\subsection{Data Reduction}
We followed the standard data reduction procedures, including bias
subtraction, overscan correction, and trimming and flat-fielding
using IRAF/CCDRED \citep{ste87}. We used the twilight sky flat
fields obtained before and after the observations for flat-fielding.
In order to increase the signal-to-noise ratio and to reach
low limiting magnitudes, we took multiple exposures of each
region. These images were aligned and median-combined into a master
image for each region using images of both the long and short
exposures. For the photometry calibration, we obtained calibration
images for each chip of the CFH12K. We used 20--40 isolated bright
stars to derive the point-spread function (PSF) of each calibration
image, and we performed PSF photometry \citep{ste87} using
IRAF/DAOPHOT to obtain the instrumental magnitudes.

For standard star transformation in each band, we observed the SA 98
region \citep{lan92} on the same night. The standard star
observations were carefully designed to ensure that the maximum 
number of stars were imaged in each chip of the CFH12K mosaic CCD. The
number of observed standard stars was 8--19 for each chip. We
derived the standard magnitudes of the programme stars by using the
following equations given on the CFHT web page:
\begin{equation}
    V = v-0.12(X-1)+0.014(V-I)+z_{v}
\end{equation}
\begin{equation}
    I = i-0.04(X-1)+0.050(V-I)+z_{i}
\end{equation}
where $V$ and $I$ are the standard magnitudes and $v$ and $i$ are
the instrumental magnitudes. The terms $z_{v}$ and $z_{i}$ are the
zero points of the photometry that were determined from the
photometry of the standard stars in SA 98, and $X$ is the airmass.
Fig. 3 shows the residuals of the standard star transformation in
the $V$ and $I$ bands.

\begin{figure}
 \vspace{-14pt}
 \epsfig{figure=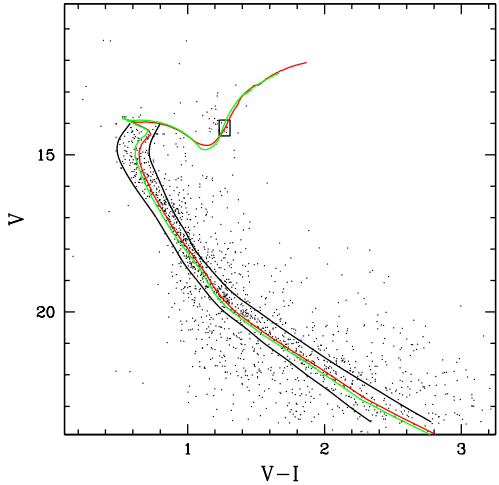, height=0.470\textwidth, width=0.470\textwidth}
 \vspace{-5mm}
 \caption{CMD of the central region ($r < 5^{\prime}$) of NGC 1245 with 
isochrones fitted by eyeball (red line) and $\chi^2$ minimum (green line).
Black lines indicate the boundaries of main-sequence stars and giant box.
}
 \vspace{0mm}
 \label{iso_fit1}
\end{figure}

\begin{figure}
 \vspace{-14pt}
 \epsfig{figure=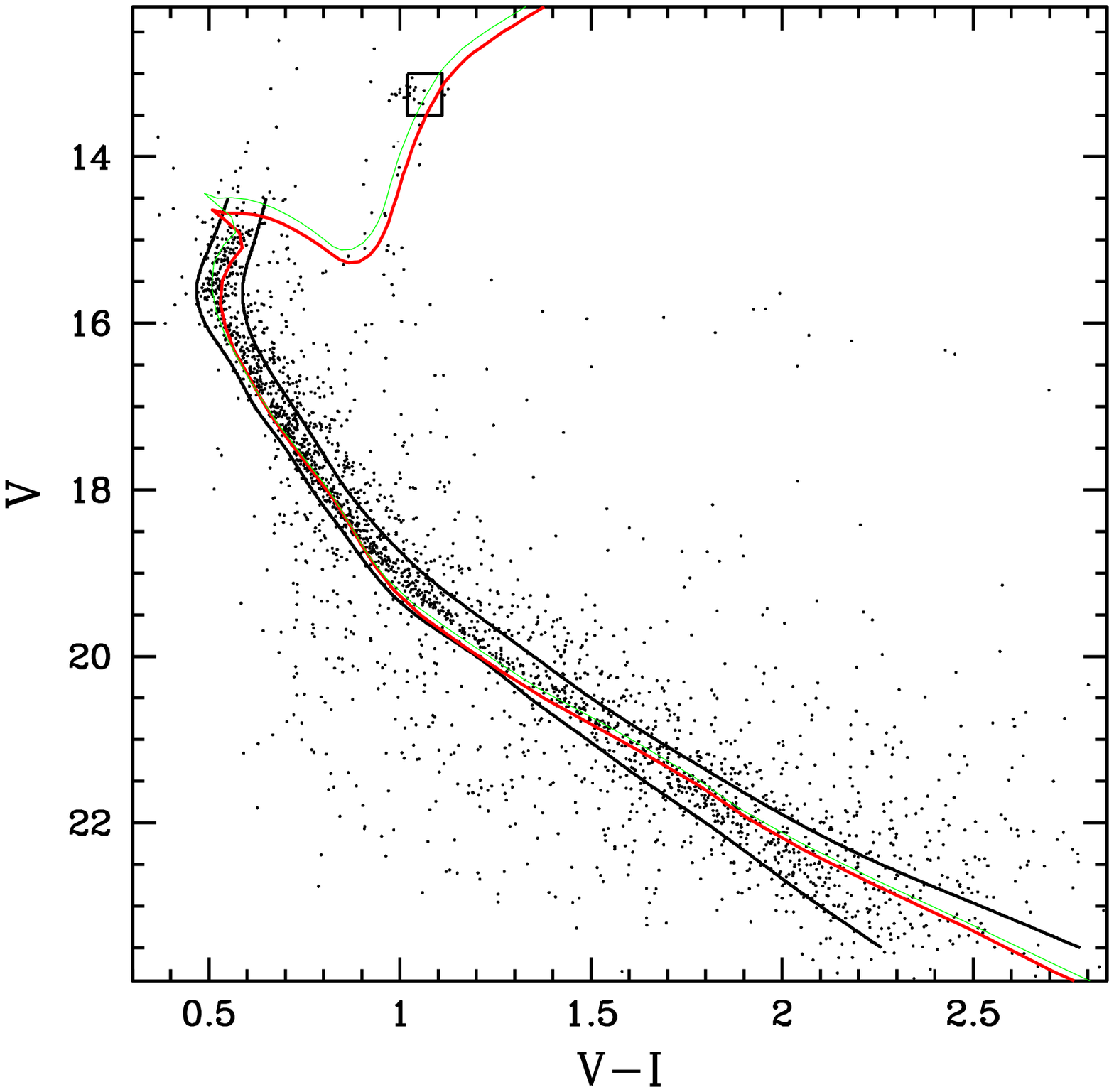, height=0.470\textwidth, width=0.470\textwidth}
 \vspace{-5mm}
 \caption{CMD of the central region ($r < 5^{\prime}$) of NGC 2506 with 
isochrones fitted by eyeball (red line) and $\chi^2$ minimum (green line).
Black lines indicate the boundaries of main-sequence stars and giant box.
}
 \vspace{-2mm}
 \label{iso_fit2}
\end{figure}

\begin{figure}
 \vspace{-1pt}
 \epsfig{figure=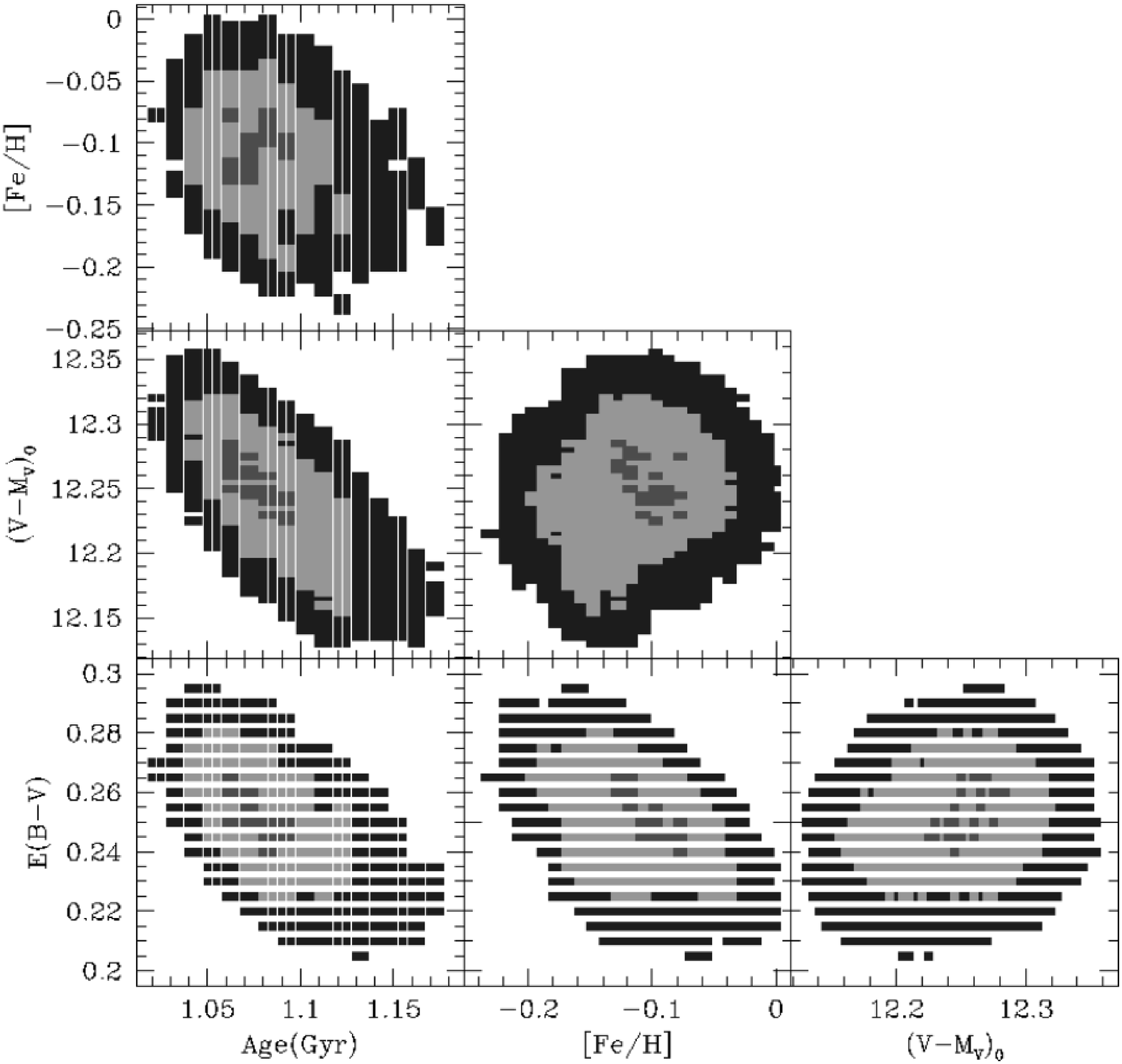, height=0.420\textwidth, width=0.420\textwidth}
 \vspace{2mm}
 \caption{Confidence regions of the joint variation in age, metallicity, 
distance modulus and reddening for NGC 1245. 
The three colors red, green, and blue represent the confidence limits 
of 2, 4, 6 $\sigma$, respectively, corresponding to $\Delta \chi^2 = 4, 16, 36$ }
 \vspace{0mm}
 \label{iso_fit}
\end{figure}

\begin{figure}
 \vspace{3pt}
 \epsfig{figure=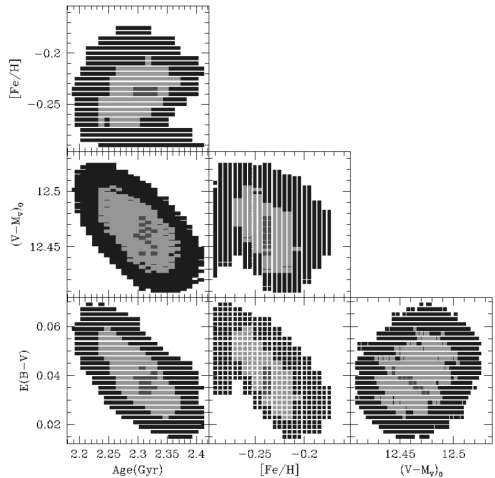, height=0.42\textwidth, width=0.42\textwidth}
 \vspace{2mm}
 \caption{Confidence regions of the joint variation in age, metallicity, 
distance modulus and reddening for NGC 2506. 
The three colors red, green, and blue represent the confidence limits 
of 2, 4, 6 $\sigma$, respectively, corresponding to $\Delta \chi^2 = 4, 16, 36$ }
 \vspace{-2mm}
 \label{iso_fit}
\end{figure}

To correct the incompleteness of the photometry, we generated
artificial stars on the CCD frames using the addstar task in the
IRAF/DAOPHOT utility and performed reduction in the same manner
as the observed frames. The additional artificial stars were randomly
generated in each unit-magnitude bin on all original images. To
prevent crowding in the images, the number of artificial stars were
controlled to be less than 10\% of the detected stars, and we
repeated this for 40 trials. The incompleteness factor ($\Lambda$)
of the photometric data was determined in the central and outer
regions of each cluster. Finally, the incompleteness correction
factor was obtained from $\Lambda = N_{r}/N_{a}$, where $N_{a}$ is
the number of added stars and $N_{r}$ is the number of recovered
stars. Fig. 4 shows the relationship between $\Lambda$ and $V$
magnitude.

\section{Physical Parameters}

We determined the basic physical parameters of the two open clusters 
by analyzing the color magnitude diagrams (CMDs). 
We used Yonsei-Yale ($Y^2$) isochrones \citep{dem04} to fit to the 
observed CMDs. We assumed $E(V-I)/E(B-V) = 1.25$ and
a total-to-selective extinction ratio $R_{V}$ of 3.2 \citep{car89}. 
Initially, we adjusted the distance, age, metallicity, and interstellar 
reddening by eyeball fitting (Fig. 5 and Fig. 6) but we determined the best
fit parameters by applying $\chi^2$ minimization algorithm as described 
below.

In order to reduce the effects of field star contamination, we constructed
cluster CMDs using stars in the central region ($R < 5^{\prime}$) of the
cluster. We also defined the boundaries of main-sequence stars,
which are the blue and red edge along main-sequence stars. These boundaries
are tightly selected in order to exclude binary sequence. According to
evolution tracks, the blue edge of giant clump stars are more massive and
more evolved than the red edge of the giant clump. And the giant branch of
$Y^2$ isochrone represents the red edge of giant stars. So we defined a region
of red edge of the giant clump. In Fig. 5 and Fig. 6, we showed the boundaries,
and the red edge regions. The $Y^2$ isochrones provide an interpolation scheme
to calculate isochrones for an arbitrary age and metallicity.

Following \citet{bur04}, for a given set of fitting parameters (age,
metallicity, distance modulus and interstellar reddening), we defined 
the $\chi^2$ as 

\begin{equation}
    \chi^2 = \sum_{i} \chi_i^2
\end{equation}
where the sum is over all stars selected for isochrone fitting and 
$\chi_i^2$ is the contribution from star $i$ which is defined as  
\begin{equation}
    \chi_i^2 \equiv [V_{m} - V_{o}]_i^2 + [(V-I)_{m} - (V-I)_{o}]_i^2.
\end{equation}
Here, $V$ and $V-I)$ are magnitude and color of stars and the subscripts 
$m$ and $o$ stand for model and observation, respectively. We assumed 
an equal weight for all stars used in the calculation of $\chi^2$ 
in order to avoid overemphasizing the main-sequence turn-off where the 
systematic difference between theoretical isochrones are more pronounced 
\citep{gro03, bur04} although their photometric errors are the smallest. 
We performed a grid search over age, metallicity, distance modulus and 
reddening to find the best-fit parameters which minimize the $\chi^2$.
In table 3, we presents the ranges and intervals of parameters used for 
NGC 1245 and NGC 2506. 

The confidence limits on the derived parameters are obtained by scaling the 
resulting $\chi^2$ statistic \citep{bur04}
\begin{equation}
    \Delta \chi^2 = \frac{\chi^2 -\chi_{min}^2}{\chi_{min}^2/\nu}
\end{equation}
where $\chi^2_{min}$ is the minimum $\chi^2$ and $\nu$ is the number 
of degrees of freedom. In Fig. 7 and Fig. 8, we plotted the confidence regions 
for the joint variation in the isochrone parameters representing
the $\Delta \chi^2 = 4, 16, 36$ (red, green, blue), which correspond
to 2, 4, and 6 $\sigma$ error respectively. As shown in Fig. 7 and
Fig. 8, the local minimum of $\Delta \chi^2$ is fairly well defined and 
the $\Delta \chi^2$ surface can be approximated by a paraboloid near minimum.
In this approximation, the statistical errors (1 $\sigma$ errors) are the
projection of the $\Delta \chi^2=1.0$ extent of the paraboloid on the 
parameter axes. The best-fit parameters and their statistical 
errors ($\sigma_{stat}$) are given in Table 4.
As can be seen in Fig. 7 and Fig. 8, the statistical errors in the best-fit
parameters are smaller in NGC 2506 than NGC 1245. The larger errors in
NGC 1245 is caused by the broad main-sequence near the turn-off in NGC 1245.

\begin{table*}
 \centering
 \begin{minipage}{70mm}
 \raggedright
  \caption{Isochrone grid parameters for NGC 1245 and NGC 2506}
   \begin{tabular}{@{}ccccc@{}} \hline
    Cluster      & \multicolumn{2}{c}{NGC 1245} &  \multicolumn{2}{c}{NGC 2506}\\ 
                    &         Range           &  Interval     &      Range                  &      Interval    \\ \hline\hline
    age(Gyr)   & $1.02 \sim 1.18$    &    0.005      &    $2.19 \sim 2.41$     &  0.005    \\ 
 
    [Fe/H]      & $ -0.24 \sim 0.0 $   &  0.005       &    $-0.29 \sim -0.16$  &  0.005   \\
$(V-M_V)_0$  & $12.13 \sim 12.36$  &  0.005       & $12.41 \sim 12.53$    &  0.002  \\
  $E(B-V)$     & $0.2 \sim 0.3$        &  0.005     & $0.015 \sim 0.070$     &  0.002  \\

   \hline
  \end{tabular}
\\
 \vspace{-0mm}
 \end{minipage}
\end{table*}

\begin{table*}
 \centering
 \begin{minipage}{145mm}
 \raggedright
  \caption{Physical parameters of NGC 1245 and NGC 2506}
   \begin{tabular}{@{}cccccc@{}} \hline
    Cluster      &                           &         $E(B-V)$        &     $(V - M_V )_0 $     &       age(Gyr)           &    [Fe/H]     \\ \hline\hline
    NGC 1245  & best $\chi^2 _{min}$ &        $0.24  $        &         $12.25 $             &          $1.08 $           &      $-0.08$                     \\
                    &$zero point+\sigma_{z}$    &        0.30              &      12.20                    &      1.06                   &       -0.12    \\
                    &$zero point-\sigma_{z}$    &        0.21              &      12.32                    &      1.08                   &       -0.09    \\
                    &  $R_V=3.0$            &         0.26              &       12.33                   &      1.06                    &       -0.12    \\
                    &  $R_V=3.4$            &        0.25                &      12.18                    &      1.06                   &       -0.11    \\
                    &  $V_{lim}=18.0$     &        0.27                &      12.33                    &       0.9                   &       -0.09    \\
                    &  $V_{lim}=21.0$     &        0.225              &      12.23                    &      1.08                   &       0.00    \\
                    &  Padova (Eyeball)  &        0.12                &      12.45                  &      1.00                   &       0.07     \\
                    &  $Y^2$ (Eyeball)      & $0.22 $               &  $12.35       $             &          $1.0$              &     $0.0 $                     \\
                   & $\sigma_{stat}$ & $\pm 0.01 $         &       $\pm 0.02$        &        $\pm 0.01$     &      $\pm 0.02 $                \\
                   & $\sigma_{sys}$ & $\pm 0.05 $         &       $\pm 0.12$        &        $\pm 0.09$     &      $\pm 0.06 $                \\
                   & $\sigma_{tot}$   & $\pm 0.05 $         &       $\pm 0.12$        &        $\pm 0.09$     &      $\pm 0.06 $                \\
    &  Previous            & $0.26^2, 0.28^8, 0.28^3$&  $13.2^2, 12.0^3, 12.4^{5} $ & $ 0.8^2 , 1.1^3  $        &  $-0.04^3,   -0.05^{6}$ \\
    &                            & $ 0.29^{5}, 0.21^6   $    &  $ 12.27^{6}, 12.1^8 $         & $ 0.9^{5} , 1.05^{6} $  &  $ +0.06^8 , +0.14^9$    \\
  NGC 2506  & best $\chi^2 _{min}$&         $0.035$           &           $12.47$            &         $ 2.31 $          &         $ -0.24 $                  \\
                    &$zero point+\sigma_{z}$   &        0.08              &      12.38                    &      2.30                   &       -0.24    \\
                    &$zero point-\sigma_{z}$   &         0.00              &      12.53                    &      2.32                   &       -0.24    \\
                    &  $R_V=3.0$           &          0.035             &           12.47                &          2.33                &           -0.24   \\
                    &  $R_V=3.4$           &          0.035             &           12.46                &           2.32                &          -0.24   \\
                    &  $V_{lim}=18.0$    &          0.02               &           12.33                &          2.48                &           -0.22   \\
                    &  $V_{lim}=21.0$    &          0.03               &           12.32                &          2.47                &          -0.23   \\
                    &  Padova (Eyeball)  &          0.02             &           12.4                  &           2.5                 &            -0.32   \\ 
                    &  $Y^2$ (Eyeball)     &          $0.02 $       &  $12.4          $             &          $2.19$             & $-0.2  $                     \\
                   & $\sigma_{stat}$ & $\pm 0.004 $       &        $\pm 0.01$       &         $\pm 0.01$     &      $\pm 0.01 $           \\
                   & $\sigma_{sys}$ & $\pm 0.04 $         &       $\pm 0.08$        &        $\pm 0.16$     &      $\pm 0.06 $                \\
                   & $\sigma_{tot}$   & $\pm 0.04 $         &       $\pm 0.08$        &        $\pm 0.16$     &      $\pm 0.06 $                \\
   &  Previous & $0.1^1, 0 \sim 0.07^4 $& $11.7^1, 12.6^4, 12.95^{7}$&$1.6 \sim 2.0^4, 1.8^{7}$&$-0.32 \sim 0.07^4, -0.32^{7}$\\
   &                  & $0.10^8 ,          $          & $11.8^{8}  $                       & $   $                                 &$ -0.41^8$                       \\
   \hline
  \end{tabular}
\\
$^1$ \citet{pur64},   $^2$ \citet{car94},       $^3$ \citet{wee96},     $^4$ \citet{mar97},
$^5$ \citet{sub03},  $^6$ \citet{bur04},  $^7$ \citet{hen07} ,  $^8$ \citet{jan79},  $^9$ \citet{lyn87}

 \vspace{2mm}
 \end{minipage}
\end{table*}
The statistical errors presented in Table 4 are very small.
However, the uncertainties of the physical parameters 
can be larger than these errors because they merely represent 
the statistical errors associated with the $\chi^2$ minimization.
There are other sources of errors that affect the accuracy 
of the derived physical parameters systematically. 
Possible causes of systematic errors are the uncertainties related to 
the photometric calibration, assumed parameters such as the 
total-to-selective extinction ratio $R_V$, limiting magnitudes of 
observed data $V_{lim}$ and the theoretical models for isochrone fittings.
In order to know the systematic errors in the derived physical parameters, 
we performed some experiments by varying the assumed parameters and  
theoretical models. We present the results in Table 4.

From equation (1) and (2), systematic errors associated with the calibration
of the photometry can be caused by uncertainties in the extinction and 
transformation coefficients and zero points. But we examined only the effect
of the uncertainties in the zero points ($\sigma_{z}(V)$ and $\sigma_{z}$)
because the calibration coefficients which are provided by CHFT web site
are assumed to be well determined. We derived physical parameters from the
data with the zero point shifts of $\pm \sigma_z(V) = 0.05$ and 
$\pm \sigma_z(V-I) = 0.10$ and found that the systematic errors in $E(B-V)$
and $(V-M_V)_0$ are relatively large.

Since systematic uncertainties in $R_V$ affect the cluster parameters,
variations in $R_V$ alter the $(V-I)$ color at fixed $A_V$ \citep{bur04},
we performed $\chi^2$ minimization procedures to derive cluster parameters 
with $R_V = 3.0$ and $R_V = 3.4$ using the same grid search. We found that 
the systematic errors associated with the $R_V$ variation are similar to or
less than the statistical errors, except for those in the distance modulus 
of NGC 1245.

In Fig. 5 and Fig. 6, there is a non-negligible discrepancy between the
observations and the theoretical isochrones for the lower main-sequence 
stars ($V > 19$). This effect is most serious for stars in $ 19 < V < 21$ of
NGC 2506. Due to this discrepancy, the derived cluster parameters can be 
changed if we assume different limiting magnitude for isochrone fittings.
To estimate the uncertainties related to the discrepancy between the models
and the observations, we performed the same procedures again to derive
the cluster parameters with limiting magnitudes of $V_{lim} = 18.0$ and
$V_{lim} = 21.0$. The systematic errors related to the $V_{lim}$ are generally
larger than the statistical errors and they are larger in NGC 1245. But, 
the systematic errors in $(V-M_V)_0$ are larger in NGC 2506. The reason for
the larger uncertainties in the parameters of NGC 1245 is partly due to the 
broad distribution of stars near the turn-off and upper main-sequence of 
NGC 1245. The large systematic error in $(V-M_V)_0$ in NGC 2506 seems to be
caused by the maximum discrepancy between the models and observations at
$ 19 < V < 21$ in NGC 2506.

Since the cluster parameters derived from isochrone fittings significantly
depend on the theoretical models, we used the Padova isochrones \citep{gir00}
for comparison models. However, we did not apply $\chi^2$ minimization
procedures because we do not have interpolation programs for the Padova 
isochrones. Rather, we applied eye-ball fittings for both the $Y^2$ isochrones
and the Padova isochrones and compared the derived cluster parameters. As can
be seen in Table 4, the systematic errors resulting from the theoretical 
models are much larger than the statistical errors but comparable to the
systematic errors associated with the assumed parameters. In general, the 
systematic errors associated with the theoretical models are larger in
NCC 1245 than NGC 2506. The largest difference is observed in the age of 
NGC 2506.

We calculated the systematic errors of cluster parameters from the derived
values in Table 4 by assuming that we explored typical values of $R_V$,
$V_{lim}$ and zero point shifts in the derivation of cluster parameters. 
Actually, we considered the standard deviations of the cluster parameters
resulting from different sets of assumed parameters (zero points, $R_V$ 
and $V_{lim}$) and theoretical models as the systematic errors of the 
cluster parameters. Then, the total systematic error ($\sigma_{sys}$) can  
be determined by a quadratic sum of the systematic errors as
\begin{equation}
{\sigma_{sys}}^2={\sigma_z}^2+{\sigma(R_V)}^2+{\sigma(V_{lim})}^2+{\sigma(iso)}^2
\end{equation}
here $\sigma(R_V)$ and $\sigma(V_{lim})$ are systematic errors associated with
the parameters $R_V$ and $V_{lim}$, respectively, and $\sigma(iso)$ is the
systematic error related to the theoretical models. Then, the total 
error ($\sigma_{tot}$) can be written as
\begin{equation}
{\sigma_{tot}}^2={\sigma_{stat}}^2+{\sigma_{sys}}^2
\end{equation}
where $\sigma_{stat}$ is the statistical error related to the $\chi^{2}$
minimization. In Table 4, we present the total uncertainties in each cluster
along with the statistical errors and the systematic errors.  
However, the actual uncertainties in the derived cluster parameters 
can be larger than these values since we did not consider the errors 
associated with the unresolved binaries and the field star contamination. 
Our estimates of the systematic errors are similar to those of \citet{bur04}
who derived the cluster parameters by the same method as ours 
and the total uncertainties are comparable to those of \citet{jan11}
who derived the cluster parameters using synthetic CMDs.

\section{Luminosity Function}

\subsection{Construction of Luminosity Function}

Our observed area was wide enough to cover the field region as well
as the cluster region. To derive the LFs of the open clusters, we
first determined the cluster extent $R_{cl}$ by examining the
surface density distribution of the observed area. The radius at
which the surface number density radial profiles of the stars
becomes constant was taken to be $R_{cl}$. The field regions can
then be defined as the region $R > 2R_{cl}$. From this, we derived
the field star population, which was required to correct the open
cluster LFs. The cluster radii were $R_{cl} = 16^{\prime}$ and
$19^{\prime}$ for NGC 1245 and NGC 2506, respectively. We derived
the LFs by counting the number of stars in each bin of magnitude
$\Delta m=1$ for the $V$-band photometry of the cluster region with
the correction of field stars as follows:
\begin{equation}
  N = \frac {N_{cl}}{\Lambda} - \frac{N_{f}}{\Lambda} \frac{A_{cl}}{A_{f}}
\end{equation}
where $N_{cl}$ is the number of stars for each magnitude bin in the
cluster region, $N_{f}$ is that of the field region, and $A_{cl}$
and $A_{f}$ are the areas of the cluster and field regions,
respectively. 
Here, we assumed that all the field regions have the same
LF, but this is not usually true because there are fluctuations in
the surface number densities of the field stars as well as
inhomogeneous field star populations. 

\begin{figure}
 \epsfig{figure=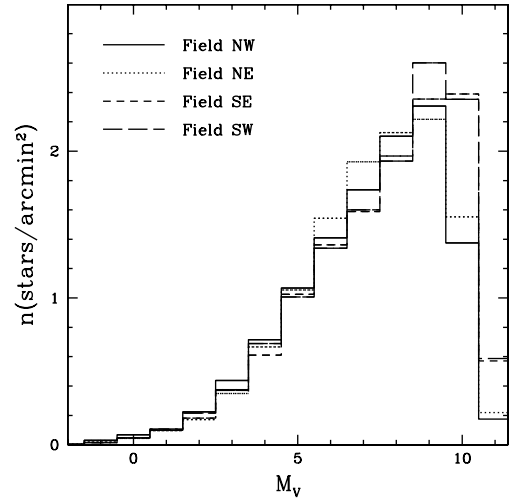, height=0.480\textwidth, width=0.480\textwidth}
 \vspace{-5mm}
 \caption{
Luminosity functions of four field regions around NGC 1245.}
 \label{iso_fit}
 \vspace{4mm}
\end{figure}

\begin{figure}
 \epsfig{figure=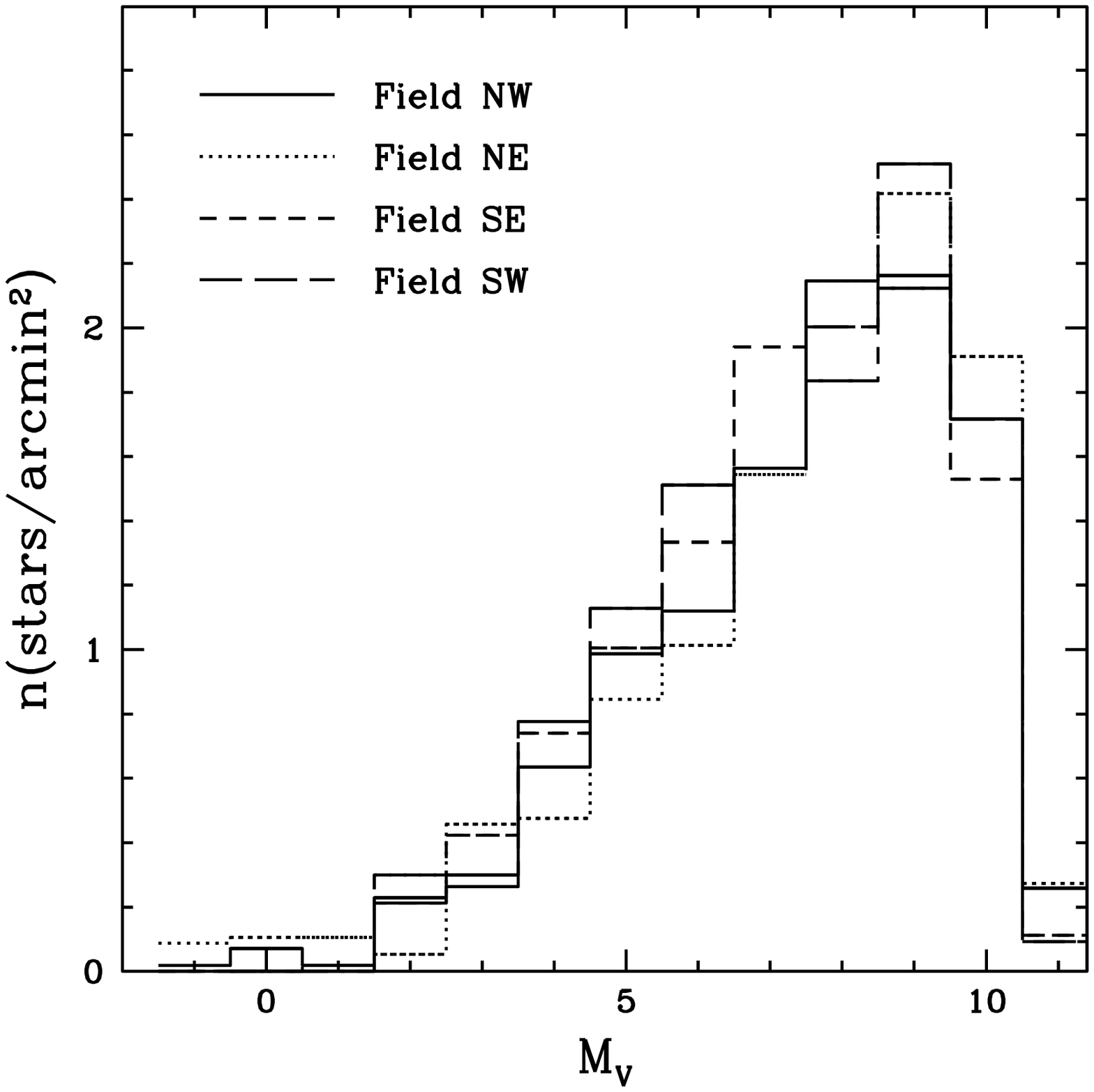, height=0.480\textwidth, width=0.480\textwidth}
 \vspace{-5mm}
 \caption{
Luminosity functions of four field regions around NGC 2506. }
 \label{iso_fit}
 \vspace{4mm}
\end{figure}

\begin{table}
 \centering
 \begin{minipage}{53mm}
 \vspace{0mm}
  \raggedright
  \caption{Dispersions in the LFs of field regions of NGC 1245 and NGC 2506}
   \begin{tabular}{@{}ccc@{}} \hline
                &    \multicolumn{2}{c}{Dispersion (stars/(arcmin)$^2$)}  \\ 
      $M_V$    & NGC 1245    &  NGC 2506  \\ \hline\hline
        -1.0   &    0.007    &    0.036   \\
         0.0   &    0.010    &    0.046   \\
         1.0   &    0.004    &    0.041   \\
         2.0   &    0.022    &    0.090   \\
         3.0   &    0.033    &    0.081   \\
         4.0   &    0.039    &    0.117   \\
         5.0   &    0.024    &    0.100   \\
         6.0   &    0.079    &    0.193   \\
         7.0   &    0.137    &    0.166   \\
         8.0   &    0.083    &    0.128   \\
         9.0   &    0.142    &    0.164   \\
        10.0   &    0.459    &    0.135   \\
        11.0   &    0.192    &    0.082   \\
   \hline
  \end{tabular}
 \end{minipage}
 \vspace{2mm}
\end{table}


In order to take into account the uncertainties arising from the 
fluctuations in different
directions, we constructed LFs of the field regions surrounding the clusters 
(see Fig. 9 and Fig. 10 for NGC 1245 and NGC 2506, respectively). 
The LFs of the NGC 1245 field regions were very similar to each other except 
for the large difference at $M_{V}\sim10$. 
This large difference was caused 
partly by the different limiting magnitudes of the individually observed regions
and partly by large photometric errors at the faintest magnitudes.
While the four field LFs were nearly the same in NGC 1245, the field
LFs in the south of NGC 2506 were much different from those in the
north of NGC 2506. 
More specifically, the number of stars in the
southern fields of NGC 2506 was larger than that in the northern
fields. 
Since NGC 2506 is located in the north of the Galactic
plane, the difference seen in Fig. 10 was due to the density gradient
along the Galactic latitude.
Because the LFs of the open clusters were affected by those of the field regions, the dispersions
in the field LFs are listed in Table 5 as a measure of the uncertainties caused by the
correction.                                                                            
If we consider the errors of the cluster LFs as the dispersions 
in the field LFs scaled to the area of the cluster, they become larger than 
the Poisson errors, but this seems to be more a meaningful way
to represent the errors in the cluster LFs \citep{phe93}.

\begin{figure}
 \vspace{0mm}
 \epsfig{figure=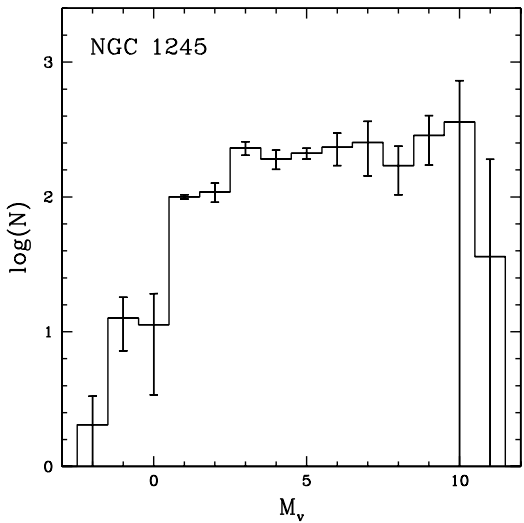, height=0.50\textwidth, width=0.50\textwidth}
 \vspace{-5mm}
 \caption{
Luminosity function of NGC 1245. Error bars represent the logarithmic
dispersion of field LFs scaled to the cluster area.
}
 \vspace{0mm}
 \label{iso_fit}
\end{figure}

\begin{figure}
 \vspace{0mm}
 \epsfig{figure=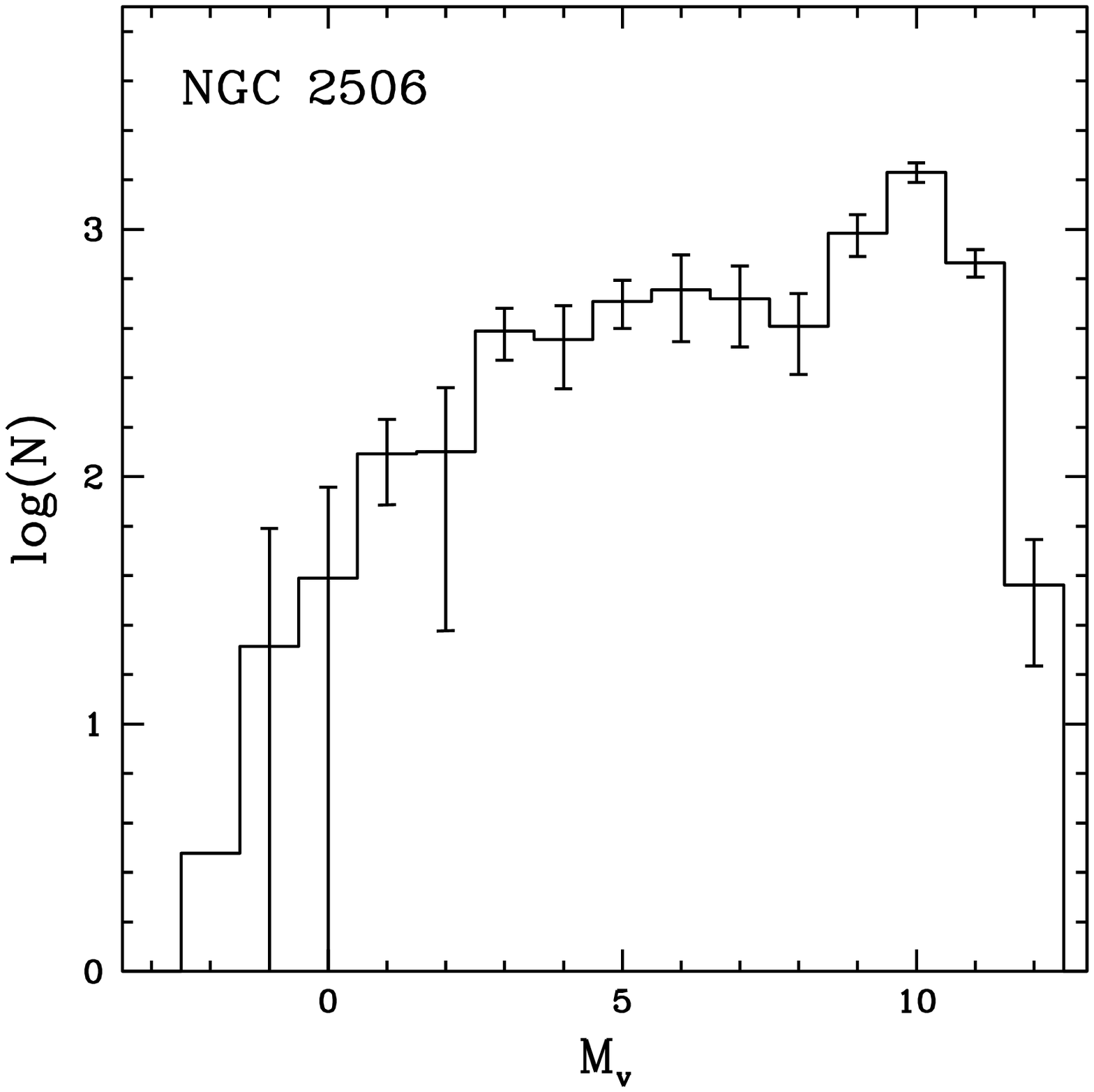, height=0.50\textwidth, width=0.50\textwidth}
 \vspace{-5mm}
 \caption{
Luminosity function of NGC 2506. Error bars represent the
logarithmic dispersion of field LFs scaled to the cluster area. }
 \vspace{-2mm}
 \label{iso_fit}
\end{figure}

\begin{figure}
 \epsfig{figure=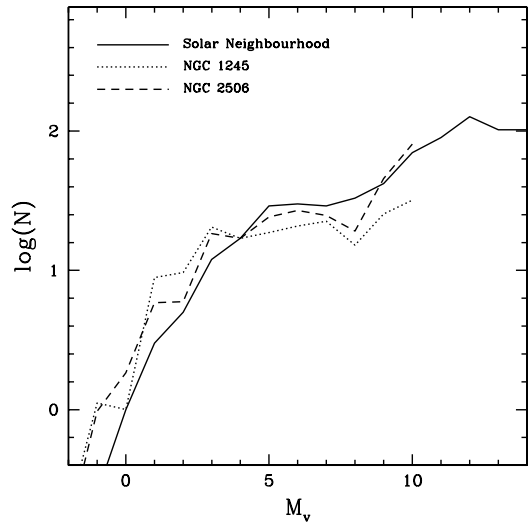, height=0.50\textwidth, width=0.50\textwidth}
 \vspace{-5mm}
 \caption{
LFs compared with that of the solar neighbourhood. All LFs have been
normalized to the value at $M_V$ =4.0 }
 \vspace{2mm}
 \label{iso_fit}
\end{figure}

\subsection{Characteristics of Luminosity Functions}

The LFs of NGC 1245 and NGC 2506 are shown in Fig. 11 and Fig.
12. The logarithmic error bar was scaled by area using Table 5 as
follows:
\begin{equation}
  e_{+} = log(N + \Delta \times A_{cl}) - log(N) \\
\end{equation}
\begin{equation}
  e_{-} = log(N) - log(N - \Delta \times A_{cl}) \\
\end{equation}
Here, $e_{+}$ and $e_{-}$ are the positive and negative errors,
respectively, and $\Delta$ is the LF dispersion of the field region
as listed in Table 5. Our observing limit was $M_V \approx 10$, but
near $M_V \approx 10$ for NGC 1245, the uncertainties increased
dramatically because of the different limiting magnitudes of each
observing region.

Stars brighter than $M_V \approx 1$ are thought to be turn-off and
giant stars in NGC 1245 and NGC 2506. Comparing the bright parts of
the LFs of NGC 1245 and NGC 2506, the number of bright stars in NGC
1245 is larger than that in NGC 2506. The reason for the larger number
of brighter stars in NGC 1245 is the younger age of NGC 1245. The age of
NGC 2506 is nearly twice that of NGC 1245. And the giant clump stars
of $M_V \approx 0 $ are more developed in NGC 2506.

The LF of NGC 1245 is nearly flat. The flat LF of NGC 1245 could be caused by
the dynamical evolution which makes the low-mass stars evaporate from the
cluster because the LF of the cluster is expected to increase monotonically
toward the faint magnitudes to at least $M_V \approx 10$. The LF of NGC 2506
is similar to that of NGC 1245, but they do differ. While the LF of NGC 1245
is relatively flat between $M_V=3$ and $M_V=10$, the LF of NGC 2506 shows
a gradual increase toward fainter magnitudes. It seems unlikely that this
difference is due to the errors in the field star correction  because the
dispersions in the field LFs are not large enough to erase the rising
tendency toward the faint magnitudes. It seems worthy to note that the LF of
NGC 2506 is derived from a slightly larger area than that of NGC 1245,
which may include regions where evaporating stars are likely to be located,
at least temporarily. The radial dependencies of the LFs of NGC 1245 and
NGC 2506 show that the peak at $M_V=10$ in the LF of NGC 2506 is most
pronounced in the LF derived from the outer regions.

In Fig. 13, we compare the LFs of NGC 1245 and NGC 2506 with that of
the solar neighbourhood \citep{bin98}. 
Because turn off magnitudes are about $M_V \approx 2.0$ and 3.5 for NGC 1245 
and NGC 2506, respectively, 
we normalized the LFs at $M_V=4.0$ to avoid evolutionary effects.
In this figure, we can see a 
similar global pattern between the solar neighbourhood and the open
clusters, but the number of bright stars in the LFs of NGC 1245 and
NGC 2506 is slightly larger than that of the solar neighbourhood,
while the opposite is true for the number of faint stars. We believe
this feature is caused by the evaporation of low-mass stars in NGC
1245 and NGC 2506 because they are old enough to expect this
kind of dynamical evolution and mass segregation.

\begin{figure}
 \vspace{1mm}
 \epsfig{figure=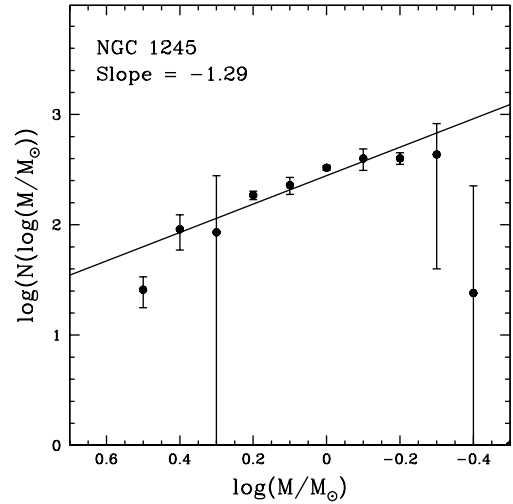, height=0.480\textwidth, width=0.480\textwidth}
 \caption{
Mass function of NGC 1245. Error bars represent the logarithmic dispersion of field MFs scaled to the cluster area.
The fitting range is $log(M/M_{\odot}) = -0.2 \sim 0.3$.
}
 \label{iso_fit}
 \vspace{2mm}
\end{figure}

\begin{figure}
 \vspace{1mm}
 \epsfig{figure=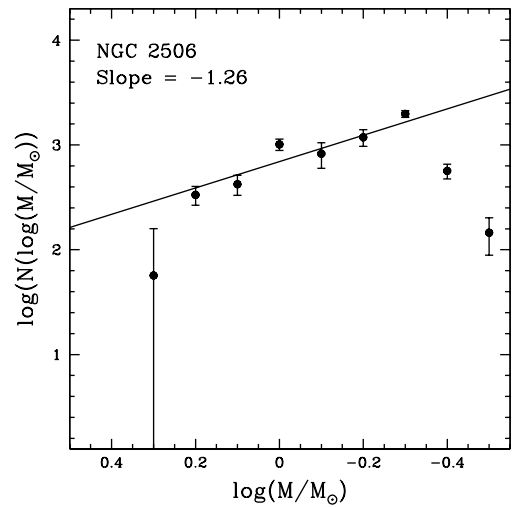, height=0.480\textwidth, width=0.480\textwidth}
 \caption{
Mass function of NGC 2506. Error bars represent the logarithmic dispersion of field MFs scaled to the cluster area.
The fitting range is $log(M/M_{\odot}) = -0.2 \sim 0.1$.
}
 \label{iso_fit}
 \vspace{2mm}
\end{figure}

\section{Mass Function} 

The mass function is a useful tool for studying the IMF and to
understand the star formation process in general as well as the
structure and evolution of a stellar system. In order to construct
the mass function of NGC 1245 and NGC 2506, we used the
mass-luminosity relation of \citet{sch92} for stars with $M_V < 7.0$
and that of \citet{hen93} for $M_V = 7$--$17.6$.  Fig. 14 and Fig.
15 show the mass functions of NGC 1245 and NGC 2506 derived by
counting the stars in the mass interval $\Delta log(M/M_{\odot}) =
0.1$ inside $R_{cl}$. The uncertainties in the mass function were
derived from the dispersions of the field region mass functions,
similar to the uncertainties of the LFs. We derived the slope of
mass function
\begin{equation}
  \Gamma = \frac {d log N(log (M/M_{\odot}))}{d log(M/M_{\odot})}.
\end{equation}
by a least-squares fit, weighted by dispersions in the MFs of field regions.
We obtained $\Gamma = -1.35 \pm 0.04$ for NGC 1245 and $\Gamma = -1.26 \pm 0.07$
for NGC 2506 with fitting ranges of $log(M/M_{\odot}) = -0.2 \sim 0.3$
and $log(M/M_{\odot}) = -0.2 \sim 0.1$, respectively for NGC 1245 and NGC 2506.

The derived $\Gamma$ values of the two
clusters were slightly shallower than the slope of the solar
neighbourhood IMF: $\Gamma = -1.35$ \citep{sal55}. In the case of NGC
1245, the present $\Gamma$ value is similar to those reported by
\citet{car94} and \citet{sub03} who derived a similar but shallower
slope than the Salpeter value. The slopes of the mass functions of
NGC 1245 and NGC 2506 were thought not to be the same as the slope of
the respective IMFs because of their age; evaporation of the
low-mass stars results in a shallower slope.

\section{Summary} 

We  conducted deep and wide $VI$ photometry of old open clusters
NGC 1245 and NGC 2506 based on CCD observations using the CFHT.
The present photometry is suitable for investigating dynamical evolution
in open clusters because it is deep enough to reach down to 
$M_{V}\approx10$ and wide enough to cover the entire regions of the
two clusters including the field regions for an effective correction of the
field star contamination.

We derived the physical parameters of the two clusters using detailed
isochrone fittings based on $\chi^2$ minimization: $E(B-V) = 0.24 \pm0.05$ and
$ 0.03 \pm0.04$, $(V-M_V)_0  = 12.25 \pm0.12$ and $12.47 \pm0.08$,
$age(Gyr) = 1.08 \pm0.09$ and $2.31 \pm0.16$, and [Fe/H]= $-0.08 \pm0.06$
and $-0.24 \pm0.06$ for NGC 1245 and NGC 2506, respectively.
The quoted errors are the total errors which are quadratic sum of statistical 
errors associated with the $\chi^2_{min}$ and systematic errors related to
the assumed parameters and theoretical models.
The present estimates of the physical parameters of two clusters are in
good agreement with the previous estimates.

We also derived the LFs of NGC 1245 and NGC 2506. 
The LF of NGC 1245 shows a flat profile between $M_V=3$ and $M_V=10$, 
whereas the LF of NGC 2506 displays a slight rise. 
This difference seems to be not due to the errors related to the field star 
correction but due to the
dynamical structures, which can differ because dynamical evolution
can be different for clusters of a similar age as it is affected by
the environment as well as by the internal properties of the cluster.
We derived the mass functions of NGC 1245 and NGC 2506, which show
slightly shallower slopes than that of the solar neighbourhood IMF.
The shallower slope is understandable if we consider that the
derived mass function is the present day mass function, which is
thought to be different from the IMF of the clusters due to
dynamical evolution, since the ages of NGC 1245 and NGC 2506 are old
enough for mass segregation and evaporation of low-mass stars.
We will discuss the dynamical properties, structures and 
halo of these two clusters in a forthcoming paper. 

\section*{Acknowledgments}
This work was supported for two years by Pusan National University 
Research Grant. YWK was supported in part by the research  
grant from KASI (Korea Astronomy and Space Science Institute).

\clearpage

\end{document}